\documentclass[submission,copyright,creativecommons]{eptcs}
\providecommand{\event}{GANDALF 2016} 
\usepackage{breakurl}             
\usepackage{xspace}
\usepackage{proof}
\usepackage{xcolor}
\usepackage{amsmath}
\usepackage{stmaryrd}
\usepackage{leftidx}
\usepackage{cmll}
\usepackage{amssymb}
\usepackage{graphicx} 
\usepackage{rotating}

\usepackage{listings} 
\usepackage{qtree}
\usepackage{subcaption}

\colorlet{lprolog}{blue!70!black}

\newcommand{\fpc}{FPC\xspace}

\newcommand{\pcert}{ProofCert\xspace}

\newcommand\proofsystem[1]{\mbox{\slshape #1}\xspace}

\newcommand\LKF  {\proofsystem{LKF}}
\newcommand\aLKF {\hbox{\proofsystem{LKF}\kern-2pt$^a$}\xspace}

\newcommand\aLJF {\hbox{\proofsystem{LJF}\kern-2pt$^a$}\xspace}

\newcommand{\async}[2]{\vdash#1\mathbin{\Uparrow} #2}

\newcommand{\Async}[3]{#1\async{#2}{#3}}  
\newcommand{\Sync }[3]{#1\sync{#2}{#3}}  

\newcommand{\wedgep}{\wedge^{\!+}}
\newcommand{\wedgen}{\wedge^{\!-}}
\newcommand{\veep}{\vee^{\!+}}
\newcommand{\veen}{\vee^{\!-}}

\newcommand{\andClerk}[3]{{\wedge_c}(#1,#2,#3)}
\newcommand{\falseClerk}[2]{f_c(#1,#2)}
\newcommand{\orClerk}[2]{{\vee_c}(#1,#2)}
\newcommand{\allClerk}[2]{\forall_c(#1,#2)}

\newcommand{\storeClerk}[3]{\hbox{\sl store}_c(#1,#2,#3)}

\newcommand{\trueExpert }[1]{{\true_e}(#1)}
\newcommand{\andExpert}[3]{{\wedge_e}(#1,#2,#3)}
\newcommand{\andExpertLJF}[6]{{\wedge_e}(#1,#2,#3,#4,#5,#6)}
\newcommand{\orExpert  }[3]{{\vee_e}(#1,#2,#3)}
\newcommand{\someExpert}[3]{\exists_e(#1,#2,#3)}

\newcommand{\initExpert}[2]{\hbox{\sl init}E(#1,#2)}
\newcommand{\cutExpert}[4]{\hbox{\sl cut}_e(#1,#2,#3,#4)}
\newcommand{\decideExpert}[3]{\hbox{\sl decide}_e(#1,#2,#3)}
\newcommand{\releaseExpert}[2]{\hbox{\sl release}_e(#1,#2)}

\newcommand{\blue}[1]{{\color[rgb]{0,0,1} #1}}
\newcommand{\tupp}[2]{\blue{\langle #1,}#2{\blue{\rangle}}}

\newcommand{\mkpos}[1]{\partial\kern -1pt_{\scriptscriptstyle +}\kern -1pt(#1)}
\newcommand{\mkneg}[1]{\partial\kern -1pt_{\scriptscriptstyle -}\kern -1pt(#1)}

\newcommand{\delayop}{\ensuremath{\partial}}

\newcommand{\tr}[2]{\lbrack #1 \rbrack_{#2}}

\newcommand{\str}[2]{ST_{#2}(#1)}

\newcommand{\trlab}[1]{\lbrack #1 \rbrack}
\newcommand{\trseq}[2]{\lbrack #1 \rbrack_{#2}}

\newcommand{\rel}{R}

\newcommand{\wld}{W}
\newcommand{\val}{V}
\newcommand{\m}{\mathcal{M}}
\newcommand{\prop}{\mathcal{P}}

\newcommand{\relfo}{R}

\newcommand{\init}{init}

\newcommand{\lwedge}{L\wedge}
\newcommand{\rwedge}{R\wedge}
\newcommand{\lvee}{L\vee}
\newcommand{\rvee}{R\vee}
\newcommand{\lbox}{L\square}
\newcommand{\rbox}{R\square}
\newcommand{\ldiamond}{L\lozenge}
\newcommand{\rdiamond}{R\lozenge}

\newcommand{\impl}{\supset}

\newcommand{\logick}{K}

\newcommand\labk {\proofsystem{G3K}}

\newcommand{\delp}[1]{{#1}^{\delayop^{+}}}

\usepackage{microtype}
\newtheorem{theorem}{Theorem}
\title{Certification of Prefixed Tableau Proofs for Modal Logic}
\author{
Tomer Libal \qquad\qquad Marco Volpe
\institute{Inria and LIX, \'Ecole Polytechnique\\
France}
\email{\quad tomer.libal@inria.fr \quad\qquad marco.volpe@inria.fr}
}
\def\titlerunning{Certification of Prefixed Tableau Proofs for Modal Logic}
\def\authorrunning{T. Libal \& M. Volpe}
\begin{document}
\maketitle
\begin{abstract}
	Different theorem provers tend to produce proof objects in different formats and this is especially the case for modal logics, where several deductive formalisms (and provers based on them) have been presented.
	This work falls within the general project of establishing a common specification language in order to certify proofs given in a wide range of deductive formalisms.
	In particular, by using a translation from the modal language into a first-order polarized language and a checker whose small kernel is based on a classical focused sequent calculus, we are able to certify modal proofs given in labeled sequent calculi, prefixed tableaux and free-variable prefixed tableaux.
	We describe the general method for the logic K, present its implementation in a Prolog-like language, provide some examples and discuss how to extend the approach to other normal modal logics.
\end{abstract}

\section{Introduction}

Modal logics are very popular and feature in many areas of computer science,
including formal verification, knowledge representation, the field of logics of programs,
computational linguistics and agent-based systems.
Two common approaches for the automatic proving of modal theorems are the tableau method \cite{fitting1972tableau}
and the resolution principle  \cite{Ohlbach1988}. Theorem provers based on such approaches normally contain non-trivial
optimizations and cores which might compromise the amount of trust we can place in them. Nevertheless, only
few of these provers do actually return an evidence supporting their results and even these evidences might not be checkable by
a computer.

\pcert\ \cite{erc} is a project whose main goal is the certification of a wide range of
proof evidences. By using well-established concepts of proof theory,
\pcert\ proposes \emph{foundational proof certificates} (\fpc) as a framework
to specify proof evidence formats. Describing a format in terms of an \fpc
allows software to check proofs in this format, much like a context-free
grammar allows a parser to check the syntactical correctness of a program. The
parser in this case would be a kernel: a small and trusted component that checks
a proof evidence with respect to an \fpc specification.

\textsf{Checkers} \cite{chihaniLR15} is a generic proof certifier based on the \pcert\ ideas.
It allows for the certification of arbitrary proof evidences using various trusted kernels.
The certification is carried out by using dedicated \fpc specifications which guide the construction
of proofs in the target kernels.
A particularly trusted and low-level kernel is the focused classical sequent calculus \LKF~\cite{LiaMil09}. In
\cite{MilVol15}, a translation from the language of the labeled sequent system $\labk$~\cite{Neg05} for propositional modal logic into the language of \LKF was described.
$\labk$ is of interest when trying to certify proofs of modal theorem provers
due to its close relationship with the refutational technique of prefixed tableaux, on which
many modal theorem provers are based.

In this paper, we propose two distinct \fpc specifications, both relying on such a translation. The first one requires a detailed proof description from the prover and allows for a step-by-step checking, while the second one only needs some core information about the original proof and operates by performing some proof reconstruction.
These specifications enable the automated checking of proof evidences coming from different deductive formalisms for the modal logic K. In particular, we will show how to apply them to the certification of proofs given in $\labk$, in Fitting-style prefixed tableaux and in a free-variable variant of prefixed tableau systems.
While the first calculus is designed for positive proofs, the other two are based on a refutational method.
Still, the most dramatic change in the form of the derivations is presented in the third method, as
the free-variable optimization introduces the
notion of meta-variables and can significantly change the structure of the proofs generated.
Proof evidences arising from these formalisms, when paired with the corresponding specification, can be automatically
certified by \textsf{Checkers} over the \LKF\ kernel.
We show, by means of examples, that using the \pcert\ flexible notion of a proof evidence and
\textsf{Checkers} modular design, we are able to support proof checking for these three different formalisms, by making use of the same translation.

To the best of our knowledge, the work presented here is the first attempt to independently certify the proofs generated
by propositional modal theorem provers.
The approach closest to ours is probably Dedukti's \cite{boespflug2012lambdapi}
independent certification for the classical first-order tableau prover Zenon modulo \cite{cauderlier2015checking}.

In the next section, we present some background on ProofCert, modal logic and theorem proving.
In Section~\ref{sec:cert}, we describe the different \fpc specifications. Such specifications are then used
in order to enhance the capabilities of \textsf{Checkers}, as we demonstrate on some examples. In Section~\ref{sec:conclusion}, we
conclude and discuss some possible future work.

\section{Background}
\subsection{A general proof checker}
There is no consensus about what shape should a formal proof evidence take. The notion of structural proofs, which is based
on derivations in some calculus, is of no help as long as the calculus is not fixed. One of the ideas of the \pcert\ project
is to try to amend this problem by defining the notion of a foundational proof certificate (\fpc) as a pair of an arbitrary
proof evidence and an executable specification which denotes its semantics in terms of some well known target calculus,
such as the Sequent Calculus. These two elements of an \fpc\ are then
given to a universal proof checker which, by the help of the \fpc, is capable of deriving a proof in the target calculus.
Since the proof generated is over a well known and low-level calculus which is easy to implement, one can obtain
a high degree of trust in its correctness.

The proof certifier \textsf{Checkers} is a $\lambda$Prolog \cite{Miller2012} implementation of this idea.
Its main components are the following:

\begin{itemize}
  \item {\bf Kernel.} The kernels are the implementations of several trusted proof calculi. Currently, there
    are kernels over the classical and intuitionistic focused sequent calculus. Section \ref{sec:lkf} is
  devoted to present \LKF{}, i.e.~the classical focused sequent calculus that will be used in the paper.
  \item {\bf Proof evidence.} The first component of an \fpc, a proof evidence is a $\lambda$Prolog description
    of a proof output of a theorem prover. Given the high-level declarative form of $\lambda$Prolog, the structure and
    form of the evidence are very similar to the original proof.
    We will see the precise form of the different proof evidences we handle in Section \ref{sec:cert}.
  \item {\bf \fpc specification.} The basic idea of \textsf{Checkers} is to try and generate a proof of the theorem
    of the evidence in the target kernel. In order to achieve that, the different kernels have additional predicates
    which take into account the information given in the evidence. Since the form of this information is not known
    to the kernel, \textsf{Checkers} uses \fpc specifications in order to interpret it. These
    logical specifications are written in $\lambda$Prolog and interface with the kernel in a sound way in order
    to certify proofs. Writing these specifications is the main task for supporting the different outputs of
    the modal theorem provers we consider in this paper and they are, therefore, explained in detail in Section \ref{sec:cert}.
\end{itemize}

\subsection{Classical Focused Sequent Calculus}
\label{sec:lkf}

Theorem provers usually employ efficient proof calculi with a lower degree of trust. At the same time, traditional proof calculi like the
sequent calculus enjoy a high degree of trust but are very inefficient for proof search.
In order to use the sequent calculus as the basis of automated deduction, much more structure within proofs needs to be established.
Focused sequent calculi, first introduced by Andreoli \cite{andreoli1992logic} for linear logic,
combine the higher degree of trust of sequent calculi with a more efficient proof search. They take advantage of the fact that
some of the rules are ``invertible'', i.e. can be applied without requiring backtracking, and that some other rules can ``focus''
on the same formula for a batch of deduction steps. In this paper, we will make use of the
classical focused sequent calculus (\LKF) system defined in \cite{LiaMil09}. Fig. \ref{fig:lkf} presents,
in the black font, the rules of \LKF.

Formulas in \LKF\ can have either positive or negative polarity and are constructed from atomic
formulas, whose polarity has to be assigned, and from logical connectives whose polarity is pre-assigned.
The connectives $\wedge^-,\vee^-$ and $\forall$ are of negative polarity, while $\wedge^+, \vee^+$ and $\exists$ are of positive polarity.

Deductions in \LKF\ are done during invertible or focused phases. Invertible phases correspond to the application of invertible rules
to negative formulas while a focused phase corresponds to the application of focused rules to a specific, focused, positive formula.
Phases can be changed by the application of structural rules.
A polarized formula $A$ is a \emph{bipolar formula} if $A$ is a
positive formula and no positive subformula occurrence of $A$ is in
the scope of a negative connective in $A$.
A \emph{bipole} is a pair of a negative phase below a
positive phase within \LKF: thus, bipoles are macro inference rules in
which the conclusion and the premises are $\Uparrow$-sequents with no
formulas to the right of the up-arrow.

It might be useful sometimes to delay the application of invertible rules (focused rules) on some negative formulas (positive formulas) $A$.
In order to achieve that, we define the following delaying operators
$\delayop^+(A) = \texttt{true} \wedge^+ A$ and $\delayop^-(A) = \texttt{false} \vee^- A$.
Clearly, $A,\delayop^+(A)$ and $\delayop^-(A)$
are all logically equivalent but $\delayop^+(A)$ is always considered as a positive formula and $\delayop^-(A)$ as negative.

In order to integrate the use of \fpc into the calculus, we enrich each rule of \LKF\ with proof evidences and additional predicates,
given in blue font in Fig. \ref{fig:lkf}. We call the resulted calculus $\aLKF$. $\aLKF$ extends $\LKF$ in the following way.
Each sequent now contains additional information in the form of the proof evidence $\Xi$.
At the same time, each rule is associated with a predicate (for example \renewcommand{\initExpert}[2]{{\hbox{\sl initial$_e$}(#1,#2)}}$\initExpert\Xi l$) which,
according to the proof evidence, might prevent the rule from being called or guide it by supplying such information as
the cut formula to be used.

Note that adding the \fpc\ definitions in Fig. \ref{fig:lkf} does not harm the soundness of the system but only restricts
the possible rules which can be applied at each step. Therefore, a proof obtained using \aLKF\ is also a proof in \LKF.
Since the additional predicates do not compromise the soundness of \aLKF, we allow their definition to be external to the kernel and in fact
these definitions, which are supplied by the user, are what allow \textsf{Checkers} to check arbitrary proof formats.
Section \ref{sec:cert} is mainly devoted to the definitions of these programs for the different proof formats of the modal theorem provers.

\begin{figure}[tb]
\renewcommand{\Async}[3]{\blue{#1}\vdash#2\mathbin{\Uparrow}   #3}
\renewcommand{\Sync }[3]{\blue{#1}\vdash#2\mathbin{\Downarrow} #3}
\renewcommand{\andClerk}[3]{\blue{{\hbox{andNeg$_c$}}(#1,#2,#3)}}
\renewcommand{\falseClerk}[2]{\blue{\hbox{f$_c$}(#1,#2)}}
\renewcommand{\orClerk}[2]{\blue{{\hbox{orNeg$_c$}}(#1,#2)}}
\renewcommand{\allClerk}[2]{\blue{\hbox{all$_c$}(#1,#2)}}
\renewcommand{\storeClerk}[4]{\blue{\hbox{\sl store$_c$}(#1,#2,#3,#4)}}
\renewcommand{\trueExpert }[1]{\blue{{\hbox{true$_e$}}(#1)}}
\renewcommand{\andExpert}[3]{\blue{{\hbox{andPos$_e$}}(#1,#2,#3)}}
\renewcommand{\andExpertLJF}[6]{\blue{{\hbox{andPos$_e$}}(#1,#2,#3,#4,#5,#6)}}
\renewcommand{\orExpert  }[3]{\blue{{\hbox{orPos$_e$}}(#1,#2,#3)}}
\renewcommand{\someExpert}[3]{\blue{\hbox{some$_e$}(#1,#2,#3)}}
\renewcommand{\initExpert}[2]{\blue{\hbox{\sl initial$_e$}(#1,#2)}}
\renewcommand{\cutExpert}[4]{\blue{\hbox{\sl cut$_e$}(#1,#2,#3,#4)}}
\renewcommand{\decideExpert}[3]{\blue{\hbox{\sl decide$_e$}(#1,#2,#3)}}
\renewcommand{\releaseExpert}[2]{\blue{\hbox{\sl release$_e$}(#1,#2)}}

{\sc Invertible Rules}
\[
\infer{\Async{\Xi}{\Theta}{A\wedgen B,\Gamma}}
      {\Async{\Xi'}{\Theta}{A,\Gamma} \quad
       \Async{\Xi''}{\Theta}{B,\Gamma} \quad
       \andClerk{\Xi}{\Xi'}{\Xi''}}
\]
\[
\infer{\Async{\Xi}{\Theta}{ A\veen B,\Gamma}}
      {\Async{\Xi'}{\Theta}{ A,B,\Gamma}\quad\orClerk{\Xi}{\Xi'}}
\qquad
\infer[\dag]{\Async{\Xi}{\Theta}{ \forall x.B,\Gamma}}
      {\Async{(\Xi' y)}{\Theta}{[y/x]B,\Gamma}\quad\allClerk{\Xi}{\Xi'}}
\]
{\sc Focused Rules}
\[
\infer{\Sync{\Xi}{\Theta}{B_1\wedgep B_2}}
      {\Sync{\Xi'}{\Theta}{B_1}\quad
       \Sync{\Xi''}{\Theta}{B_2}\quad
       \andExpert{\Xi}{\Xi'}{\Xi''}}
\]
\[
\infer{\Sync{\Xi}{\Theta}{B_1\veep B_2}}{\Sync{\Xi'}{\Theta}{B_i}\qquad
       \orExpert{\Xi}{\Xi'}{i}}
\qquad\qquad
\infer{\Sync{\Xi}{\Theta}{\exists x.B}}{\Sync{\Xi'}{\Theta}{[t/x]B}\quad
                  \someExpert{\Xi}{t}{\Xi'}}
\]
{\sc Identity rules}
\[
\infer[cut]{\Async{\Xi}{\Theta}{\cdot}}
           {\Async{\Xi'}{\Theta}{B}\quad
            \Async{\Xi''}{\Theta}{\neg{B}}
            \quad \cutExpert{\Xi}{\Xi'}{\Xi''}{B}}
\qquad
\infer[init]{\Sync{\Xi}{\Theta}{P_a}}
            {\tupp{l}{\neg P_a}\in\Theta\quad\initExpert{\Xi}{l}}
\]
{\sc Structural rules}
\[
\infer[\kern -1pt release]{\Sync{\Xi}{\Theta}{N}}
               {\Async{\Xi'}{\Theta}{N}\quad\releaseExpert{\Xi}{\Xi'}}
\qquad
\infer[store]{\Async{\Xi}{\Theta}{C,\Gamma}}
             {\Async{\Xi'}{\Theta, \tupp{l}{C}}{\Gamma} \quad
              \storeClerk{\Xi}{C}{l}{\Xi'}}
\]
\[
\infer[\kern -1pt decide]{\Async{\Xi}{\Theta}{\cdot}}
              {\Sync{\Xi'}{\Theta}{P}\quad
               \tupp{l}{P}\in\Theta\quad
               \decideExpert{\Xi}{l}{\Xi'}}
\]
\caption{The augmented \LKF proof system \aLKF.  The proviso $\dag$
  requires that $y$ is not free in $\blue{\Xi,}\Theta,\Gamma,B$.
  The symbol $P_a$ denotes a positive atomic formula. }
\label{fig:lkf}
\end{figure}

\subsection{Prefixed tableaux for modal logic}
\label{sec:back-prefixed}

\subsubsection{Modal logic}

The language of \emph{(propositional) modal formulas} consists of a functionally complete set of classical propositional connectives, a \emph{modal operator} $\square$ (here we will also use explicitly its dual $\lozenge$) and a denumerable set $\cal{P}$ of \emph{propositional symbols}. Along this paper, we will work with formulas in \emph{negation normal form}, i.e., such that only atoms may possibly occur negated in them. Notice that this is not a restriction, as it is always possible to convert a propositional modal formula into an equivalent formula in negation normal form. The grammar is specified as follows:
\small
  \begin{displaymath}
  A ::= \, P \, \mid \, \neg P \, \mid \, A \vee A \, \mid \, A \wedge A \, \mid \, \square A \, \mid \, \lozenge A \,
  \end{displaymath}
  \normalsize
where $P \in \cal{P}$. We say that a formula is a \emph{$\square$-formula} (\emph{$\lozenge$-formula}) if its main connective is $\square$ ($\lozenge$).
  The semantics of the modal logic $\logick$ is usually defined by means of \emph{Kripke frames}, i.e.,~pairs $\mathcal{F}=(\wld,\rel)$ where $\wld$ is a non empty set of \emph{worlds} and $\rel$ is a binary relation on $\wld$. A \emph{Kripke model} is a triple $\m=(\wld, \rel, \val)$ where $(\wld,\rel)$ is a Kripke frame and $\val: \wld \rightarrow 2^\prop$ is a function that assigns to each world in $\wld$ a (possibly empty) set of propositional symbols.

\emph{Truth} of a modal formula at a point $w$ in a Kripke structure $\m=(\wld,\rel,\val)$ is the
smallest relation $\models$ satisfying:
\small
	\begin{eqnarray*}
		\m, w \models P
		& \quad \text{iff} \quad & P \in \val(w) \\
		\m, w \models \neg P
		& \quad \text{iff} \quad & P \not\in \val(w) \\
		\m, w \models A \vee B
		& \quad \text{iff} \quad & \m, w \models A
		\text{ or } \m, w \models B \\
		\m, w \models A \wedge B
		& \quad \text{iff} \quad & \m, w \models A
		\text{ and } \m, w \models B \\
		\m, w \models \square A
		& \quad \text{iff} \quad & \m, w' \models A
		\text{ for all } w' \text{ s.t. } w \rel w'\\
		\m, w \models \lozenge A
		& \quad \text{iff} \quad & \text{ there exists } w' \text{ s.t. } w \rel w'	\text{ and } \m, w' \models A.
	\end{eqnarray*}
\normalsize
 By extension, we write
$\m \models A$ when $\m,w \models A \mbox{ for all } w \in \wld$ and
 we write $\models A$ when $\m \models A$ for every Kripke structure $\m$.

\subsubsection{The standard translation from modal logic into classical logic}
\label{sec:std-translation}

The following \emph{standard translation} (see, e.g.,~\cite{BlaVBe07}) provides a bridge between propositional modal logic and first-order classical logic:
\begin{center}
\begin{tabular}{ccc@{\qquad\qquad}ccc}\small
	$\str{P}{x}$ &\small = & \small$P(x)$ &
\small	$\str{A \wedge B}{x}$ & \small= &\small $\str{A}{x} \wedge \str{B}{x}$\\
\small	$\str{\neg P}{x}$ &\small = & \small$\neg P(x)$ &
	\small$\str{\square{A}}{x}$ & \small= &\small $\forall y (\relfo(x,y) \impl \str{A}{y})$\\
\small	$\str{A \vee B}{x}$ & \small= &\small $\str{A}{x} \vee \str{B}{x}$ &
	\small$\str{\lozenge A}{x}$ &\small = &\small $\exists y (\relfo(x,y) \wedge \str{A}{y})$
\end{tabular}
\end{center}
where $x$ is a free variable denoting the world in which the formula is being evaluated. The first-order language into which modal formulas are translated is usually referred to as \emph{first-order correspondence language}~\cite{BlaVBe07} and consists of a binary predicate symbol $\relfo$ and a unary predicate symbol $P$ for each $P\in \prop$. When a modal operator is translated, a new fresh variable is introduced.
It is easy to show that for any modal formula $A$, any model $\m$ and any world $w$, we have that $\m,w \models A$ if and only if $\m \models \str{A}{x}[x\leftarrow w]$.

\subsubsection{Labeled sequent systems}

	 Several different deductive formalisms have been used for modal proof theory and theorem proving. One of the most interesting approaches has been presented in~\cite{Gab96} with the name of labeled deduction.
	 The basic idea behind labeled proof systems for modal logic is to internalize elements of the corresponding Kripke semantics (namely, the worlds of a Kripke structure and the accessibility relation between such worlds) into the syntax.
	 A concrete example of such a system is the sequent calculus $\labk$ presented in \cite{Neg05}.
	 \emph{$\labk$ formulas} are either \emph{labeled formulas} of the form $x:A$ or \emph{relational atoms} of the form $x \rel y$, where $x, y$ range over a set of variables and $A$ is a modal formula. In the following, we will use $\varphi, \psi$ to denote $\labk$ formulas.
	\emph{$\labk$ sequents} have the form $\Gamma \vdash \Delta$, where $\Gamma$ and $\Delta$ are multisets containing labeled formulas and relational atoms.
	In Fig.~\ref{fig:labk}, we present the rules of $\labk$, which is proved to be sound and complete for the basic modal logic $\logick$~\cite{Neg05}.
		\begin{figure}[t]\small
		{\sc Classical rules}
		\[
		\infer[\init]{{x:P,\Gamma}\vdash{\Delta, x:P}}{}
		\qquad
		\infer[\lwedge]{x:A\wedge B, \Gamma \vdash \Delta}{x:A, x:B, \Gamma \vdash \Delta}
		\quad
		\infer[\rwedge]{\Gamma \vdash \Delta, x:A \wedge B}{\Gamma \vdash \Delta, x:A & \Gamma \vdash \Delta, x:B}
		\]
		\[
		\infer[\lvee]{x:A\vee B, \Gamma \vdash \Delta}{x:A,\Gamma \vdash \Delta & x:B,\Gamma \vdash \Delta}
		\quad
		\infer[\rvee]{\Gamma \vdash \Delta, x:A \vee B}{\Gamma \vdash \Delta, x:A, x:B}
		\]
		{\sc Modal rules}
		\[
		\infer[\lbox]{x:\square A, x\rel y,\Gamma \vdash \Delta}{y:A,x:\square A,x \rel y, \Gamma \vdash \Delta}
		\quad
		\infer[\rbox]{\Gamma \vdash \Delta, x:\square A}{x \rel y, \Gamma \vdash \Delta, y:A}
\quad
		\infer[\ldiamond]{x:\lozenge A, \Gamma \vdash \Delta}{x \rel y, y:A,\Gamma \vdash \Delta}
		\quad
		\infer[\rdiamond]{x\rel y, \Gamma \vdash \Delta, x:\lozenge A}{x \rel y, \Gamma \vdash \Delta, x:\lozenge A, y:A}
		\]
		In $\rbox$ and $\ldiamond$, $y$ does not occur in the conclusion.
		\caption{$\labk$: a labeled sequent system for the modal logic $\logick$}
		\label{fig:labk}
		\end{figure}

\subsubsection{Prefixed tableau systems}
	\label{sec:fitting-tableaux}
Prefixed tableaux can also be seen as a particular kind of labeled deductive system. They were introduced in~\cite{fitting1972tableau}.
The formulation that we use here is closer to the one in~\cite{Fit07} and it is given in terms of unsigned formulas.
A \emph{prefix} is a finite sequence of positive integers (written by using dots as separators). Intuitively, prefixes denote possible worlds and they are such that if $\sigma$ is a prefix, then $\sigma.1$ and $\sigma.2$ denote two worlds accessible from $\sigma$. A \emph{prefixed formula} is $\sigma:A$, where $\sigma$ is a prefix and $A$ is a modal formula in negation normal form.
A prefixed tableau proof of $A$ starts with a root node containing $1:A$, informally asserting that $A$ is false in the world named by the prefix $1$. It continues by using the branch extension rules given in Figure~\ref{fig:fit-rules}. We say that a branch of a tableau is a \emph{closed branch} if it contains $\sigma:P$ and $\sigma:\neg P$ for some $\sigma$ and some $P$. The goal is to produce a \emph{closed tableau}, i.e.,~a tableau such that all its branches are closed.
		\begin{figure}[tb]\small
		{\sc Classical rules}
		\[
		\infer[\wedge_F]{\sigma: A,\,\sigma:B}{\sigma: A \wedge B}
		\qquad
		\infer[\vee_F]{\sigma: A \quad \mid \quad \sigma:B}{\sigma: A \vee B}
		\]
		{\sc Modal rules}
		\[
		\infer[\square_F]{\sigma.n:A}{\sigma:\square A}
		\qquad
		\infer[\lozenge_F]{\sigma.n:A}{\sigma:\lozenge A}
		\]
		In $\square_F$, $\sigma.n$ is used. In $\lozenge_F$, $\sigma.n$ is new.
		\caption{A prefixed tableau system for the modal logic $\logick$}
		\label{fig:fit-rules}
		\end{figure}
Classical rules in Figure~\ref{fig:fit-rules} are the prefixed version of the standard ones. For what concerns the modal rules, the $\lozenge$ rule applied to a formula $\sigma:A$ intuitively allows for generating a new world, accessible from $\sigma$, where $A$ holds, while the $\square$ rule applied to a formula $\square:A$
allows for moving the formula $A$ to an already existing world accessible from $\sigma$. We say that a prefix is \emph{used} on a branch if it already occurs in the tableau branch and it is \emph{new} otherwise.

\subsubsection{Free-variable prefixed tableau systems}
\label{sec:fv-tableaux}

Prefixed tableau systems have a deficiency that is also common in first-order sequent calculi. Resolution methods \cite{Robinson1965},
which introduce meta-variables and unification may have an exponential speed-up in proof complexity over sequent calculi \cite{BaazL92}.
In a similar way, free-variable prefixed tableaux \cite{reeves1987semantic} aim at improving prefixed tableaux by the introduction of meta variables and simple unification.
This construct allows for the delaying of the $\Box_F$ rule and might result with shorter proofs, as can be seen in Fig. \ref{fig:pt} where
the free-variable tableau has 9 rule applications versus the 12 of the standard tableau proof.
The addition of meta-variables comes with the cost that careful restrictions must be posed on the tableau proofs in order to preserve soundness.
In particular, the unification of these meta-variables must be restricted in order to prevent such unsoundness.
In this paper, we are not interested in proof generation but in the structure of proofs only and will therefore omit further discussion
on this topic. The interested reader can refer to \cite{Beckert97a} for further reading.

\section{Certification of tableau modal proofs}
\label{sec:cert}

	\subsection{A translation from the modal language into a first-order polarized language}
	\label{sec:translation}

	In~\cite{MilVol15}, it has been shown how it is possible to translate a modal formula $A$ into a polarized first-order formula $A'$ in such a way that a strict correspondence between rule applications in a G3K proof of $A$ and bipoles in an \LKF proof of $A'$ holds. Such a correspondence has been used in order to prove some adequacy theorem and to define a focused version of G3K. Here we will further exploit it for checking labeled sequent and prefixed tableaux derivations in the augmented variant \aLKF.

	The translation is obtained from the standard translation of Section~\ref{sec:std-translation} by adding some elements of polarization. First of all, when translating a modal formula into a polarized one, we are often in a situation where we are interested in putting a delay in front of the formula only in the case when it is negative and not a literal. For that purpose, we define $\small\delp{A}$, where $A$ is a modal formula in negation normal form, to be $A$ if $A$ is a literal or a positive formula and $\delayop^+(A)$ otherwise.

	%
%
	%
	Given a world $x$, we define the translation $\tr{.}{x}$ from modal formulas in negation normal form into polarized first-order formulas as:
	\begin{center}
	\begin{tabular}{ccc@{\qquad\qquad}ccc}
		\small$\tr{P}{x}$ &\small = &\small $P(x)$ &
		\small$\tr{{A} \wedge {B}}{x}$ &\small = &\small $\delp{\tr{{A}}{x}} \wedgen \delp{\tr{{B}}{x}}$\\
		\small$\tr{\neg P}{x}$ &\small = &\small $\neg P(x)$ &
		\small$\tr{{A} \vee {B}}{x}$ &\small = &\small $\delp{\tr{{A}}{x}} \veen \delp{\tr{{B}}{x}}$\\
		\small	$\tr{\square {A}}{x}$ & \small= &\small $\forall y (\neg \relfo(x,y) \veen \delp{\tr{{A}}{y}})$
		 &
		\small$\tr{\lozenge {A}}{x}$ &\small = &\small $\exists y (\relfo(x,y) \wedgep \delayop^-(\delp{\tr{{A}}{y}}))$
	\end{tabular}
	\end{center}

	In this translation, delays are used to ensure that only one connective is processed along a given bipole, e.g.,~when we decide on (the translation of) a $\lozenge$-formula $\tr{\lozenge {A}}{x}$, the (translation of the) formula $A$ is delayed in such a way that it gets necessarily stored at the end of the bipole.
	Based on that, we define the translation $\trlab{.}$ from labeled
	formulas and relational atoms into polarized first-order formulas as
	$\trlab{x:A} = \tr{{A}}{x}$ and $\trlab{x\rel y} =\relfo(x,y)$.
	We will sometimes use the extension of this notion to multisets of labeled formulas, i.e.,~$\trlab{\Gamma}=\{\trlab{\varphi} \mid \varphi\in\Gamma\}$.
	Note that predicates of the form $P(x)$ and $\relfo(x,y)$ are considered as having positive polarity.
	Finally, we define a translation from $\labk$ sequents into $\LKF$ sequents:
	$$
	 \small\trseq{(\varphi_1, \ldots, \varphi_n \vdash \psi_1, \ldots, \psi_m)} \, = \, \async{\delp{\trlab{\neg\varphi_1}}, \ldots, \delp{\trlab{\neg\varphi_n}},\delp{\trlab{\psi_1}}, \ldots, \delp{\trlab{\psi_m}}}{\cdot}
	$$
	where $\trlab{\neg \varphi}$ is $\tr{{(\neg A)}}{x}$ if $\varphi =
	x:A$ and is $\neg \relfo(x,y)$ if $\varphi = x \rel y$.

	We recall here a result from~\cite{MilVol15}, where a more formal statement and a detailed proof can be found.

		\begin{theorem}\label{th:labeled-focused}
			Let $\Pi$ be a G3K derivation of a sequent $S$ from the sequents $S_1, \ldots, S_n$.
			Then there exists an \LKF derivation $\Pi'$ of $[S]$ from $[S_1], \ldots, [S_n]$, such that each rule application in $\Pi$ corresponds to a bipole in $\Pi'$. The viceversa, for first-order formulas that are translation of modal formulas, also holds.
		\end{theorem}
	Such a result is easily extended to the case of prefixed tableaux, by relying on the correspondence between prefixed tableaux and nested sequents~\cite{Fit12}, which are a subclass of labeled sequents. We omit the details.


	\subsection{Foundational proof certificate specifications}

The translation presented in Section~\ref{sec:translation} can be used in order to check labeled sequent and prefixed tableau proofs in $\LKF$.
	In fact, given the correspondence between rule applications in the original calculus and bipoles in \LKF, we can state an easy and faithful encoding of proofs, mainly based on specifying on which formulas we decide every time we start a new bipole.
We propose here two different FPC specifications.
The first one requires a quite detailed proof evidence from the prover, while in the second one we only require the prover to provide some core information about the proof evidence and we check that it is correct by reconstructing the rest of the proof.

By observing G3K and prefixed tableau rules (Section~\ref{sec:back-prefixed}), one can notice that a proof in these formalisms is fully represented by specifying:
\begin{enumerate}
	\item at each step, on which formula we apply a rule;
	\item in the case of a $\lozenge$-formula for G3K (or a $\square$-formula for tableaux), with respect to which label (prefix) we apply the rule;
	\item in the case of an initial (closure) rule, with respect to which complementary literal we apply it.
\end{enumerate}

For this reason, an adequate and detailed proof evidence of a labeled sequent or prefixed tableau proof will consist in a tree describing the original proof (we will call it a \emph{decide tree} in the following). Each node is decorated by a pair containing: (i) the formula on which a rule is applied, as explained in (1), together with (ii) a (possibly null) further index carrying additional information, to be used in cases (2) and (3) above. Formulas in the decide tree will drive the construction (bottom-up) of the \LKF derivation, in the sense that, by starting from the root, at each step, the \LKF kernel will decide on the given formula and proceed, constrained by properly defined clerks and experts, along a positive and a negative phase. Theorem~\ref{th:labeled-focused} guarantees that at the end of a bipole, we will be in a situation which is equivalent to that of the corresponding G3K or tableau proof.
As described in item (2) above, if we are applying an $\exists$-rule in \LKF, then we need further information specifying with respect to which eigenvariable we apply the rule. This is done by linking, in the proof evidence, the formula under consideration to the corresponding new-world-generating formula (a $\square$-formula in the case of G3K; a $\lozenge$-formula in the case of tableaux). Similarly, in the case of a closure (3), the additional information in the node will specify the index of the complementary literal.

In order to provide an FPC specification for a particular format, we need to define the specific items that are used to augment \LKF. In particular, the constructors for proof certificate terms and for indexes must be provided: this is done in $\lambda$-Prolog by
declaring constructors of the types \texttt{cert} and \texttt{index}. In Figure~\ref{fig:fittings-type}, we show the type declaration for the \texttt{fittings-tableaux} FPC that takes as input a decide tree corresponding to a proof, as specified above. In this declaration, we assume that \texttt{term} and \texttt{atm} are already declared, with the obvious intended meaning.

Several constructors are used to build indexes denoting formulas. \texttt{eind} is used to denote the root formula (ideally, the theorem to be proved). \texttt{lind} and \texttt{rind} are used to denote, respectively, the left and right direct subformulas of a formula (in case of a formula whose main connective is unary, we use \texttt{lind} to build the index of its only direct subformula). We have a specific constructor \texttt{bind} for subformulas of a $\square$-formula in a tableau proof (or of a $\lozenge$-formula in a G3K proof). This takes two arguments, the first one being the index of the formula itself and the second one being the index of the corresponding eigenvariable-generating formula, i.e.,~of the formula that introduced the eigenvariable used in the current rule application. Finally, \texttt{none} is used in special cases when an index is not necessary.
For instance, for the formula $((\square p) \wedge (\lozenge \neg q)) \vee (\square(\neg p \vee q))$ of Example 1 (Figure~\ref{fig:pt} (a)), we have the following (non exhaustive) mapping of indexes to formulas:

\begin{center}
\begin{tabular}{ll}
	\texttt{eind ->} $\; 1: ((\square p) \wedge (\lozenge \neg q)) \vee (\square(\neg p \vee q))$
&
	\texttt{lind(eind) ->} $\;1: (\square p) \wedge (\lozenge \neg q)$
\\

	\texttt{lind(lind(eind)) ->} $\;1: \square p$
&
	\texttt{rind(lind(eind)) ->} $\;1: \lozenge \neg q$
\\
	\texttt{bind(lind(lind(eind)),rind(lind(eind))) ->} $\;1.1: p$
\end{tabular}
\end{center}

A \texttt{dectree} represents a decide tree as described above: the two indexes taken as arguments represent, respectively, the index of the formula on which to decide and the index possibly used for additional information; the third argument (a list of \texttt{dectree}s) is used to represent the (at most two) subtrees of this formula.
Finally, we define a constructor \texttt{fitcert} that represents a certificate in the \texttt{fittings-tableaux} FPC specification. Together with the decide tree, representing the proof evidence, it takes two more arguments, which are used along the proof by the clerk and expert predicates: (i) a list of indexes to be used when storing formulas; and (ii) a list of pairs \texttt{(index, atm)} containing for each instantiated tableau $\lozenge$-formula the corresponding eigenvariable.
Both such arguments are initialized as empty when a proof checking starts.

\begin{figure}[tb]
\begin{lstlisting}[basicstyle=\scriptsize\ttfamily,breaklines=true]
	eind : index                          none : index
	lind : index -> index                 rind : index -> index
	bind : index -> index -> index

	rel : term -> term -> atm

	dectree : index -> index -> list dectree -> dectree
	fitcert : list index -> dectree -> list (pair index atm) -> cert
\end{lstlisting}
    \caption{Type declaration for the \texttt{fittings-tableaux} FPC specification.}
	\label{fig:fittings-type}
\end{figure}

In addition to the type declaration, the FPC definition
must supply the logic program defining the clerk predicates and the expert predicates.
Writing no specification for a given predicate defines that predicate to
hold for no list of arguments.
In Figure~\ref{fig:fittings-fpc}, we define clerks and experts for the \texttt{fittings-tableaux} FPC specification.
According to this specification, each decide step is completely determined by the proof evidence: in the \texttt{decide}$_e$ expert, \texttt{I} denotes the index of the formula on which to decide; notice also that the first element of the \texttt{fitcert} (the list of indexes) is reinitialized to be empty, as this list will be used, along the bipole, to construct appropriate indexes for the subformulas being created.
For example, in \texttt{orNeg}$_c$, two indexes (\texttt{lind I} and \texttt{rind I}) are created and put inside such a list. In the end of the bipole, these indexes will be used to store the subformulas with a proper index. \texttt{orNeg}$_c$ (as well as \texttt{andNeg}$_c$) is described by two cases. The first one corresponds to the case when the $\vee^-$ rule ($\wedge^-$ rule) is being applied on the formula on which we have just decided; the second one corresponds to the case when the $\vee^-$ connective ($\wedge^-$ connective) arises from the translation of a $\square$-formula ($\lozenge$-formula). In the former case, we have to take care of the indexes generated; in the latter, we do not.
With regard to \texttt{andPos}$_e$, we remark that the connective $\wedge^+$ can only occur in a formula that is the translation of a $\lozenge$-formula and for this reason we have only one case.
\texttt{release}$_c$ leaves things unchanged. In the case of \texttt{all}$_c$, we update the list of maps between formulas and corresponding eigenvariables and in \texttt{some}$_e$, by using the information contained in such a list and the information contained in the second index of the current node in the decide tree (specifying to which universal formula the current existential formula corresponds), we are able to instantiate the $\exists$ with the proper eigenvariable.
In the \texttt{store}$_c$ clerk, we use the index created so far to properly store the formula under consideration; note that in the case of relational atoms, we simply store the formula with the index \texttt{none}. Finally, in order to apply an \texttt{initial} rule, the expert \texttt{initial}$_e$ checks that the complementary formula is indexed with the second element of the decide tree (\texttt{O} in the specification).

\begin{figure}[tb]
\begin{lstlisting}[basicstyle=\scriptsize\ttfamily,breaklines=true]
$\forall$ L,I,O,D,M. decide$_e$ (fitcert L (dectree I O D) M) I (fitcert [] (dectree I O D) M).

$\forall$ I,O,H,T,M. orNeg$_c$ (fitcert [] (dectree I O [H|T]) M) (fitcert [lind I, rind I] H M).
$\forall$ E,L,D,M. orNeg$_c$ (fitcert [E|L] D M) (fitcert [E|L] D M).

$\forall$ I,O,H,G,T,M. andNeg$_c$ (fitcert [] (dectree I O [H,G|T]) M) (fitcert [lind I] H M) (fitcert [rind I] G M).
$\forall$ E,L,D,M. andNeg$_c$ (fitcert [E|L] D M) (fitcert [E|L] D M) (fitcert [E|L] D M).

$\forall$ L,I,O,T,M,F. andPos$_e$ (fitcert L (dectree I O T) M) (fitcert L (dectree I none T) M) (fitcert L (dectree I O T) M).

$\forall$ C. release$_e$ C C.

$\forall$ I,O,H,T,M. all$_c$ (fitcert [] (dectree I O [H|T]) M) (Eigen\ fitcert [lind I] H [pair I Eigen|M]).

$\forall$ I,O,H,T,M,X. some$_e$ (fitcert [] (dectree I O [H|T]) M) X (fitcert [bind I O] H M) :- (pair O X) $\in$ M.

$\forall$ C,X,Y. store$_c$ C (n (rel X Y)) none C :- !.
$\forall$ C,X,Y. store$_c$ C (p (rel X Y)) none C :- !.
$\forall$ H,T,D,M,F. store$_c$ (fitcert [H|T] D M) F H (fitcert T D M).

$\forall$ L,I,O,D,M. initial$_e$ (fitcert L (dectree I O D) M) O.
\end{lstlisting}
    \caption{Definition of the \texttt{fittings-tableaux} FPC specification.}
	\label{fig:fittings-fpc}
\end{figure}

As already remarked, the \texttt{fittings-tableaux} FPC specification allows for a very detailed and completely faithful checking of an original proof in both the G3K sequent calculus and the Fitting-style prefixed tableau system of Section~\ref{sec:fitting-tableaux}. In fact, at least as long as the logic K is considered, a proof of a modal formula $A$ in the first setting can be easily converted into a refutation of $\neg A$ in the second formalism, where: at a given connective rule in G3K corresponds the rule of the dual connective in a tableau and at an initial rule application corresponds a closure of a branch\footnote{Clearly, if one considers the two-sided sequent version of G3K given here, it is also necessary to take care of the different meaning of having a connective on the left or on the right side of the sequent. This can be done by defining a translation from two-sided G3K sequents into one-sided sequents, as shown, e.g., in~\cite{MilVol15}.}. This means that we can use the same format for proof evidence, the same translation and the same FPC specification for both the formalisms. In the case of a theorem proved in G3K, we will obtain in \LKF a proof of its translation (as defined in Section~\ref{sec:translation}); in the case of prefixed tableaux, we will obtain a proof of the translation of the negated formula. Further remarks on the implementation and on some experiments will be given in Section~\ref{sec:examples}

Having such a faithful proof representation can appear in some cases rather naive and space-consuming. It is quite common, in the context of proof checking, to work with less precise proof evidences, that only contain crucial information about a given proof, and let the checker perform some proof reconstruction of the rest.
The second FPC specification that we propose, \texttt{simpfit-tableaux} (Figures~\ref{fig:fv-type} and~\ref{fig:fv-fpc}), aims at this goal. Instead of telling LKF, at each beginning of a new bipole, on which formula to decide, we only provide the checker with the following information:
\begin{itemize}
\item a mapping between $\square$ and $\lozenge$-formulas, i.e., which eigenvariable to use at each instantiation of an existential formula;
\item a mapping between complementary literals, i.e.,~all the pairs of literals with respect to which we can apply an \texttt{initial} rule.
\end{itemize}

This information is specified in the proof evidence by two further constructors: \texttt{boxinfo} and \texttt{closure}.
Together with such a proof evidence, a \texttt{simpfit-tableaux} certificate requires other elements:
\begin{itemize}
	\item an integer, used as a flag, for dealing with the creation of indexes: $1$ means that we have just decided on a formula and thus we need to create subindexes at the first rule application concerning a connective; we have $0$ otherwise;
	\item a list of indexes, used, as in \texttt{fittings-tableaux}, for storing formulas with the proper index;
	\item a list of maps between universal formulas and eigenvariables, updated by the \texttt{all}$_c$ clerk;
	\item a further list of indexes denoting formulas on which we have already decided: since the decide step is not driven by the certificate, as in the previous FPC, this list is required in order to avoid that the checker keeps deciding on a same formula.
\end{itemize}

The \texttt{simpfit-tableaux} FPC specification is rather general and can be used to certify proofs also in formalisms that are farther from sequent calculus (and from \LKF in particular). For example, we applied this FPC to check proofs given in the free-variable tableau systems of Section~\ref{sec:fv-tableaux}. A faithful reproduction of a proof in such a formalism would not be trivial in \LKF, because free-variable tableaux admit the application, on a given branch, of a $\square$-rule before the corresponding $\lozenge$-rule (see, e.g., tableau (b) in Figure~\ref{fig:pt}).
However, we can extract from such proofs the information about eigenvariable instantiation and closures and use it while freely constructing the rest of the proof.

\begin{figure}[tb]
\begin{lstlisting}[basicstyle=\scriptsize\ttfamily,breaklines=true]
	boxinfo : index -> index -> boxinfo.
	closure : index -> index -> closure.
	use : index -> use.

	simpfitcert : int -> list index -> list closure -> list boxinfo -> list (pair index atm) -> list use -> cert.
\end{lstlisting}
    \caption{Type declaration for the \texttt{simpfit-tableaux} FPC specification.}
	\label{fig:fv-type}
\end{figure}

\begin{figure}[tb]
\begin{lstlisting}[basicstyle=\scriptsize\ttfamily,breaklines=true]
$\forall$ F,I,C,B,E,U,D,U2. decide$_e$ (simpfitcert F I C B E U) D (simpfitcert 1 [D] C B E U2) :- use D $\in$ U, U2 = U $\setminus$ {use D}.
$\forall$ F,I,C,B,E,U. decide$_e$ (simpfitcert F I C B E U) none (simpfitcert 1 [none] C B E U).

$\forall$ I,C,B,E,U. orNeg$_c$ (simpfitcert 1 [I] C B E U) (simpfitcert 0 [lind I, rind I] C B E U).
$\forall$ I,C,B,E,U. orNeg$_c$ (simpfitcert 0 I C B E U) (simpfitcert 0 I C B E U).

$\forall$ I,C,B,E,U. andNeg$_c$ (simpfitcert 1 [I] C B E U) (simpfitcert 0 [lind I] C B E U) (simpfitcert 0 [rind I] C B E U).
$\forall$ I,C,B,E,U. andNeg$_c$ (simpfitcert 0 I C B E U) (simpfitcert 0 I C B E U) (simpfitcert 0 I C B E U).

$\forall$ F,I,C,B,E,U. andPos$_e$ (simpfitcert F I C B E U) (simpfitcert 0 I C B E U) (simpfitcert 0 I C B E U).

$\forall$ C. release$_e$ C C.

$\forall$ F,I,C,B,E,U. all$_c$ (simpfitcert F [I] C B E U) (Eigen\ simpfitcert 0 [lind I] C B [pr I Eigen|E] U).

$\forall$ F,I,C,B,E,U,X. some$_e$ (simpfitcert F [I] C B E U) X (simpfitcert 0 [bind I O] C B2 E [(use I)|U]) :- (pair O X) $\in$ E, (boxinfo I O)$\in$B, B$\setminus${boxinfo I O}=B2.

$\forall$ F,I,C,B,E,U,X,Y. store$_c$ (simpfitcert F I C B E U) (n (rel X Y)) none (simpfitcert 0 I C B E U) :- !.
$\forall$ F,I,C,B,E,U,X,Y. store$_c$ (simpfitcert F I C B E U) (p (rel X Y)) none (simpfitcert 0 I C B E U) :- !.
$\forall$ F,H,T,C,B,E,U,A. store$_c$ (simpfitcert F [H|T] C B E U) (n A) H (simpfitcert 0 T C B E U) :- !.
$\forall$ F,H,T,C,B,E,U,Form. store$_c$ (simpfitcert F [H|T] C B E U) Form H (simpfitcert 0 T C B E [(use H)|U]).

$\forall$ F,I,T,C,B,E,U,Compl. initial$_e$ (simpfitcert F [I|T] C B E U) Compl :- member (closure I Compl) C.
$\forall$ F,I,T,C,B,E,U,Compl. initial$_e$ (simpfitcert F [I|T] C B E U) Compl :- member (closure Compl I) C.
$\forall$ F,I,T,C,B,E,U. initial$_e$ (simpfitcert F [I|T] C B E U) none.
\end{lstlisting}
    \caption{Definition of the \texttt{simpfit-tableaux} FPC specification.}
	\label{fig:fv-fpc}
\end{figure}

\subsection{Examples}
	\label{sec:examples}
\begin{figure}[tb]
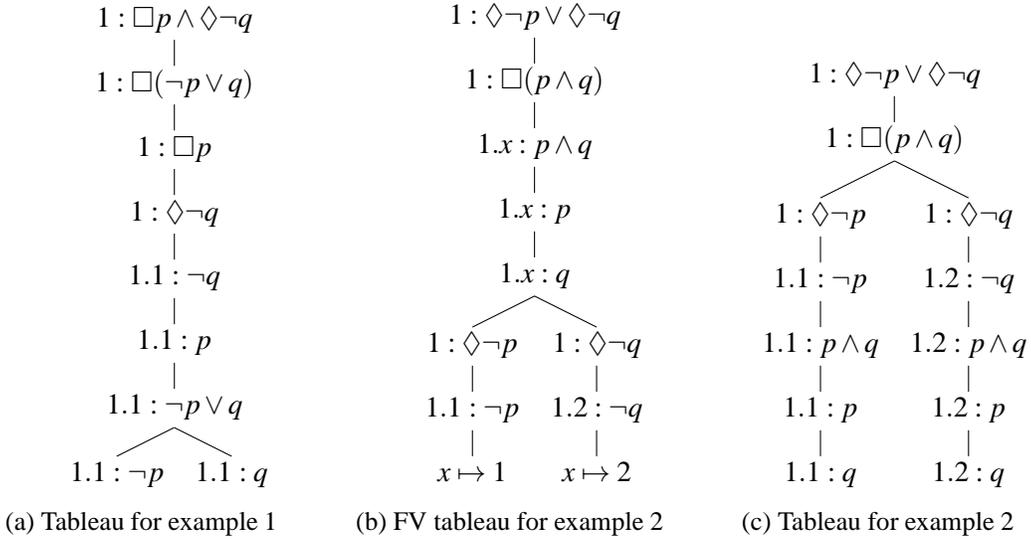

  \centering
  \begin{subfigure}[b]{0.3\textwidth}
    \centering
    \Tree [.{$1:\Box p \wedge\Diamond\neg q$} [.{$1:\Box(\neg p \vee q)$} [.{$1:\Box p$} [.{$1:\Diamond \neg q$} [.{$1.1:\neg q$} [.{$1.1:p$} [.{$1.1:\neg p\vee q$} [.{$1.1:\neg p$} ] [.{$1.1: q$} ] ] ] ] ]  ] ] ]
    \caption{Tableau for example 1}
    \label{subfig:insvariant}
  \end{subfigure}
\begin{subfigure}[b]{0.3\textwidth}
    \centering
    \Tree [.{$1:\Diamond \neg p \vee \Diamond\neg q$} [.{$1:\Box(p \wedge q)$} [.{$1.x:p\wedge q$} [.{$1.x:p$} [.{$1.x:q$} [.{$1:\Diamond \neg p$} [.{$1.1:\neg p$} [.{$x\mapsto 1$} ] ] ] [.{$1:\Diamond\neg q$} [.{$1.2:\neg q$} [.{$x\mapsto 2$} ]  ]  ]  ] ]  ] ] ]
    \caption{FV tableau for example 2}
    \label{subfig:insvariant}
  \end{subfigure}
\begin{subfigure}[b]{0.3\textwidth}
    \centering
    \Tree [.{$1:\Diamond \neg p \vee \Diamond\neg q$} [.{$1:\Box(p \wedge q)$} [.{$1:\Diamond\neg p$} [.{$1.1:\neg p$} [.{$1.1:p\wedge q$} [.{$1.1:p$} [.{$1.1:q$} ] ] ] ] ] [.{$1:\Diamond\neg q$} [.{$1.2:\neg q$} [.{$1.2:p\wedge q$} [.{$1.2:p$} [.{$1.2:q$} ] ] ] ] ] ] ]
    \caption{Tableau for example 2}
    \label{subfig:insvariant}
  \end{subfigure}
    \caption{Prefixed and free-variable prefixed tableau derivations}
	\label{fig:pt}
\end{figure}

In this section we apply the specifications from the previous section to several examples.
Since none of the theorem provers we have experimented with produced a proof as an output,
we had to modify their source codes in order to obtain some information about their execution
states and then create the proof evidences by hand. The prover we have chosen to produce this partial information with
is ModLeanTAP \cite{Beckert97a}, a free variable modal tableau prover written in Prolog.
Here we consider the following two examples:
$$
1.\; (\Box(p \Rightarrow q)) \Rightarrow ((\Box p) \Rightarrow (\Box q)) \qquad \qquad \qquad 2. \;(\Diamond \neg p \vee \Diamond \neg q)\wedge \Box (p \wedge q)
$$
The first example is taken from the ModLeanTAP self testing benchmark \footnote{A set of problems which can be found in modleantest.pl in \url{http://formal.iti.kit.edu/beckert/modlean/}.}.
The second example is more involved since it requires the setting of two different worlds for the same box formula.
This example is taken from \cite{Beckert97a} and demonstrates the importance of an \fpc specification for the proofs generated by free-variable prefixed-tableaux
as can be seen from Fig. \ref{fig:pt}.
This figure shows the prefixed tableaux and free-variable prefixed tableaux derivations of the two examples.

We have extracted two types of proofs for the above examples using ModLeadTAP. The first type of proofs contains detailed step-by-step
evidence. This form of proof can normally be extracted from prefixed tableaux and G3K proofs.
The second type contains only the essential information: the relationship between all the boxes
and diamonds and between all literal which were used for closing a branch. Since free-variable tableau provers often use
optimization techniques which breaks the relationship between their proof structure and that of normal tableaux, most
of this information loses its importance and we require only the essential information to be included.

\textsf{Checkers} can be obtained online \footnote{The exact version can be found on the ``gandalf2016'' branch in the git repository \url{https://github.com/proofcert/checkers}.}
and can be executed by running in a bash terminal: \footnote{\textsf{Checkers} depends on the $\lambda$Prolog interpreter Teyjus (\url{http://teyjus.cs.umn.edu/})}.
\begin{verbatim}
$ ./prover.sh ftab1
\end{verbatim}
where the argument is the name of the  $\lambda$Prolog module denoting the proof evidence one wishes to check.
In this example, the name is the module of the detailed proof (to be used with the \texttt{fittings-tableaux} FPC specification) of the first problem, which can be found in Fig. \ref{fig:ftab1}.
The module for the ``essential'' one (to be used with the \texttt{simpfit-tableaux} FPC specification) can be found in Fig. \ref{fig:sftab1}. More examples
can be found in the following folder (shipped together with \textsf{Checkers}): \begin{verbatim}/src/test/tableaux\end{verbatim}

\begin{figure}[tb]
  \scriptsize
  \begin{lstlisting}[language=prolog,basicstyle=\tiny]
module ftab1.                                                       % module declaration
accumulate fittings-tableaux.                                       % fpc specification module
accumulate lkf-kernel.                                              % kernel module
modalProblem "Detailed proof of ModLeanTAP problem t1"              % problem description
(((dia (-- p1)) !! (box (++ q1))) !! (dia ((++ p1) && (-- q1))))    % modal theorem
(fitcert [] (                                                       % proof evidence
 (dectree eind none [
  (dectree (lind eind) none [
   (dectree (rind (lind eind)) none [
    (dectree (lind (lind eind)) (rind (lind eind)) [
     (dectree (rind eind) (rind (lind eind)) [
      (dectree (bind ((rind eind)) ((rind (lind eind)))) none [
       (dectree (lind (bind ((rind eind)) ((rind (lind eind))))) (bind ((lind (lind eind))) ((rind (lind eind)))) []),
       (dectree (lind (rind (lind eind))) (rind (bind ((rind eind)) ((rind (lind eind))))) [])])])])])])])) [] ).
  \end{lstlisting}
\caption{src/test/tableaux/ftab1.mod}
\label{fig:ftab1}
\end{figure}

\begin{figure}[tb]
  \scriptsize
  \begin{lstlisting}[language=prolog,basicstyle=\tiny]
module sftab1.
accumulate simpfit-tableaux.
accumulate lkf-kernel.
modalProblem "Essential proof of ModLeanTAP problem t1"
(((dia (-- p1)) !! (box (++ q1))) !! (dia ((++ p1) && (-- q1))))
(simpfitcert 1 [eind]
 [ closure (lind (bind (rind eind) (rind (lind eind)))) (bind (lind (lind eind)) (rind (lind eind))),
   closure (lind (rind (lind eind))) (rind (bind (rind eind) (rind (lind eind)))) ]
 [ boxinfo (lind (lind eind)) (rind (lind eind)),
   boxinfo (rind eind) (rind (lind eind)) ] [] [] ).
  \end{lstlisting}
\caption{src/test/tableaux/sftab1.mod}
\label{fig:sftab1}
\end{figure}

\section{Conclusion}
	\label{sec:conclusion}
The examples given in the previous section were relatively simple and it is our intention to apply the tool to a real benchmark of problems, such as the one described in \cite{balsiger2000benchmark}. In order to achieve this goal, we need to find tableau provers which output detailed enough proofs.

We also aim at extending the approach to variants of the logic K. With regard to this, we remark that the translation of~\cite{MilVol15} for G3K sequents was proved to be effective not only for K but for all those modal logics characterized by Kripke frames whose relational properties can be expressed by means of geometric theories (most common modal logics fall within this class). Extending our approach to deal with such logics in the case of labeled sequent calculi is therefore straightforward. If we consider Fitting-style prefixed tableaux, however, we notice that they are typically extended to capture specific relational properties, e.g., transitivity, not by using rules that operate on the relational atoms, but rather by modifying the existing rules concerning modalities or by adding further such rules. This tends to break the strict correspondence between rule applications and bipoles, but, as observed in~\cite{MarMilVol16}, such a correspondence can be somehow recovered, e.g., by using an encoding that involves the application of a cut rule.

In the spirit of generality pursued by the ProofCert project, we also plan to consider deductive formalisms other than tableaux, e.g., resolution. Since propositional modal resolution provers are often based on a translation into a first-order language, we expect to be able to reuse part of this work also in that setting.
%

\smallskip
\noindent{\bf Acknowledgment.} This work was funded by the ERC
Advanced Grant Proof\kern 0.6pt Cert.
\bibliographystyle{eptcs}
\bibliography{main}
\end{document}